# Non-Equilibrium Orbital Transport in Terahertz Optorbitronics

**Sobhan Subhra Mishra[1,2], Ranjan Singh[3]\***

[1]*Division of Physics and Applied Physics, School of Physical and Mathematical Sciences, Nanyang Technological University, Singapore 637371*
[2]*Centre for Disruptive Photonic Technologies, The Photonics Institute, Nanyang Technological University, Singapore 639798*
[3]*Department of Electrical Engineering, University of Notre Dame, Notre Dame, IN, USA*

*\* Corresponding Author- rsingh3@nd.edu*

## Abstract

Modern information technologies rely on controlling the flow of electrons through their charge and spin. A rapidly emerging alternative is to use the orbital motion of electrons, the way they circulate around atomic sites as a new carrier of information. This orbital angular momentum (OAM) could enable more energy-efficient devices and reduce reliance on scarce heavy elements, but how orbital currents are generated and transported, especially on ultrafast timescales, remains largely unknown. In this review, we introduce *terahertz optorbitronics*, an approach that uses ultrafast femtosecond laser pulses and terahertz radiation to observe orbital transport in real time. On timescales of quadrillionth of a second, this technique allows us to track how orbital currents are launched, propagate, and convert into electrical signals in nanoscale thin-film materials. Surprisingly, recent experiments have revealed conflicting pictures such as orbital currents may travel over tens of nanometres like ballistic waves or instead decay within just a few atomic layers, highlighting a fundamental unresolved question in the field. We explain how these ultrafast measurements can disentangle orbital motion from conventional spin transport, and we highlight new materials from engineered graphene to altermagnets, that could act as tunable sources of orbital currents. We also discuss how light, electrical gating, strain, and interface design can be used to actively control orbital transport and improve its conversion into usable electronic signals.  By revealing orbital transport as a dynamic, non-equilibrium process, terahertz optorbitronics opens a new direction for nanoscale science, the one that could lead to faster, more efficient technologies operating beyond the limits of conventional spin-based electronics.

# 1. Introduction

## a. Why orbital angular momentum (OAM)

In solid-state physics, the linear combination of atomic orbitals (LCAO) serves as the foundational method for constructing electronic Bloch states. The index associated with a given Bloch state encapsulates two key aspects for a specific Bloch vector:

1. The valence orbitals of the constituent atoms
2. The intrinsic spin of the electrons.

The exploration of spin degrees of freedom and their transport properties has been central to the advancement of spintronics, a field focused on the generation, propagation, detection, and manipulation of spin information within solids. Spintronics leverages spin angular momentum (SAM) of the electrons alongside its charge for next-generation information processing and storage, offering new possibilities for device functionality and efficiency. However, the efficacy of spintronics devices faces limitations owing to the comparatively shorter relaxation length of spin, typically ranging from 1-3 nm. Furthermore, the manipulation of SAM relies on spin-orbit coupling (SOC), a property that increases with the atomic mass of the nucleus of an element. Consequently, current advancements rely on rare and uneconomic-to-extract materials such as Platinum, and Tungsten. In this regard, orbital angular momentum (OAM) provides a complementary degree of freedom to spin that can, in principle, be generated without relying on spin-orbit coupling and potentially transported over longer distances.

Historically, research on OAM transport has been limited by orbital quenching which is the suppression of orbital moment by crystal-field splitting that breaks the continuous rotational symmetry of atomic orbitals. However, recent theoretical and experimental advances have shown that OAM plays a significant role under nonequilibrium conditions, where quenching is alleviated and OAM contributes to dynamic transport processes. Moreover, unlike SAM, OAM couples directly to the phonon angular momentum (PAM) through efficient orbital–crystal-momentum ($\mathbf{L}-\mathbf{K}$) coupling, as shown in Figure 1(a). This coupling provides a pathway for angular momentum exchange between electronic and lattice degrees of freedom, enabling external control over OAM transport dynamics.

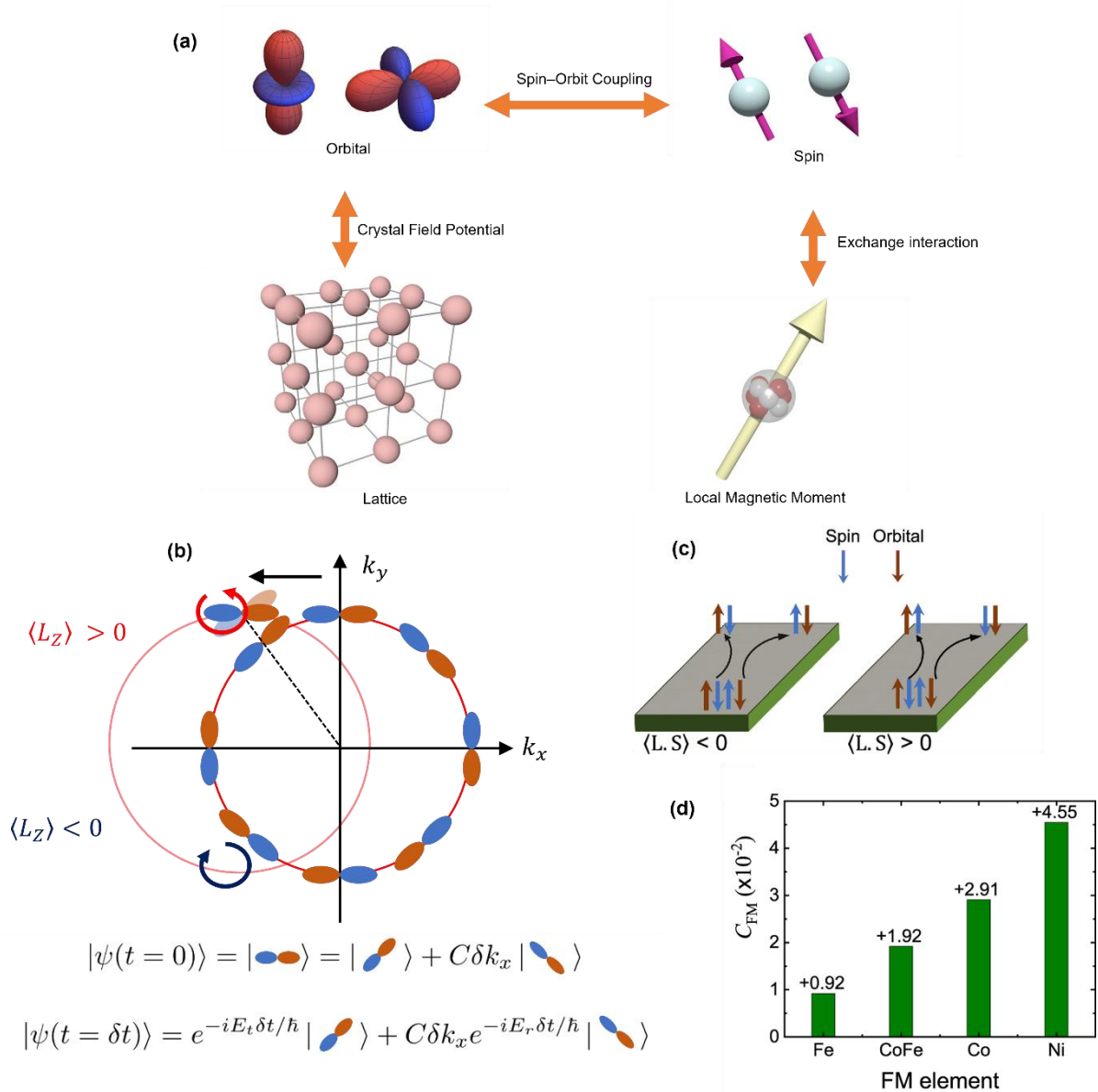

**Figure 1: Orbital Hall Effect in Solid;** (a) Coupling of angular momentum in solids, Spin angular momentum (SAM) and Local magnetic moment are coupled through various exchange interaction, SAM and Orbital angular momentum (OAM) are coupled through spin orbit coupling (SOC), OAM and lattice (Phonon angular momentum (PAM)) are coupled through crystal field potential (b) Formation of orbital texture and finite orbital angular momentum OAM along the $\boldsymbol{k} \times \boldsymbol{E}$ direction under non-equilibrium conditions, induced by a shift in $\boldsymbol{k}$ due to the applied electric field $\boldsymbol{E}$ leading to the emergence of OHE. Figure adapted from Ref[1,2] (c) The relative sign between OHE and the spin hall effect SHE determined by the $\langle \boldsymbol{L.S} \rangle$ correlation mediated by spin orbit coupling SOC. Figure adapted from Ref[1,3] (d) Magnitude of $\boldsymbol{L}$ to $\boldsymbol{S}$ conversion efficiency in various ferromagnets highlighting Co and Ni as efficient sources of orbital current. Figure adapted from[4]

## b. Charge to orbital and orbital to charge conversion: orbital Hall and inverse orbital Hall effect

Orbital currents, arising from the orbital angular momentum (OAM) of Bloch states in solids, fundamentally differ from their spin counterparts, as they derive from the orbital character of electronic states rather than spin states in vacuum. The orbital Hall effect (OHE), analogous to the spin Hall effect, can be classified into intrinsic and extrinsic mechanisms. Here, we focus on the intrinsic mechanism, wherein the ground-state electronic structure dictates orbital transport.

In systems with both inversion symmetry and time-reversal symmetry, the OAM is subject to stringent symmetry constraints:

➢ For inversion symmetric systems, $\langle L \rangle_k = \langle L \rangle_{-k}$

➢ For time reversal symmetry preserved system $\langle L \rangle_k = -\langle L \rangle_{-k}$.

The coexistence of these symmetries enforces $\langle L \rangle_k = 0$, resulting in complete suppression of OAM. Therefore, external fields are required to induce OAM in these systems.

Go et al[1,2,5]. demonstrated that an applied electric field $\boldsymbol{E}$, induces a hybridized state of radial and tangential orbital characters, generating a finite OAM proportional to $\boldsymbol{k} \times \boldsymbol{E}$ as shown in Figure 1(b) (For simplicity initial tangential orbital characteristics is assumed). This symmetry-breaking mechanism induces an OAM with opposite signs for electrons propagating in opposite directions, thereby producing a transverse orbital current to realize the OHE. Similarly, theoretical, and experimental evidence suggest that inversion symmetry breaking can induce a finite OAM from charge current due to Orbital Rashba-Edelstein Effect (OREE)[6,7]. The generation of spin current from an applied electric field typically relies on materials with strong spin-orbit coupling (SOC), such as heavy metals like platinum (Pt) or tungsten (W), where SOC facilitates the spin Hall effect (SHE). In contrast, the emergence of the orbital Hall effect (OHE) and orbital Rashba-Edelstein effect (OREE) does not require significant SOC. As a result, a finite OHE can arise even in light metals with negligible SOC, purely due to the orbital dynamics of electrons. However, when SOC is present, the spin angular momentum (SAM) and orbital angular momentum (OAM) become coupled or entangled, leading to an interplay between orbital and spin transport phenomena. This entanglement can modify the relative strengths and signs of the spin and orbital Hall effects. Specifically, the sign of the spin Hall effect relative to the orbital Hall effect is generally governed by the sign of the spin-orbit interaction term $\langle \boldsymbol{L}.\boldsymbol{S} \rangle$, as illustrated in Figure 1(c).

Following the Onsager reciprocal relation, the inverse effects of the orbital Hall effect (IOHE) and orbital Rashba-Edelstein effect (IOREE) should, in principle, enable the conversion of an orbital current into a charge current. However, the ubiquitous presence of the inverse spin Hall effect (ISHE) and inverse Rashba-Edelstein effect (IREE), and the intrinsic similarities between the symmetry and magnetization dynamics of orbital angular momentum and its spin counterpart, impose a fundamental limitation in unambiguously disentangling orbital-to-charge conversion from spin-mediated processes. Despite these similarities, spin and orbital currents exhibit inherently distinct behaviors in terms of their propagation lengths and transport dynamics, particularly on ultrafast timescales. Upon ultrafast photoexcitation in a ferromagnetic layer, ultrafast demagnetization occurs, creating a difference between equilibrium and instantaneous magnetization, resulting in spin accumulation and generation of corresponding spin current. Due to a similar magnetization-driven mechanism, an orbital accumulation may emerge, facilitated by SOC, thereby rendering ferromagnetic materials such as Ni and Co with high SOC viable sources of orbital currents as depicted in Figure 1(d).

However, while the magnetization dynamics governing spin and orbital currents exhibit striking parallels, their transport characteristics within an adjacent nonmagnetic (NM) layer can diverge significantly, depending on the thickness of the NM material. These differences arise from the distinct interfacial and bulk transport properties of spin and orbital currents particularly on ultrafast timescales. Consequently, advanced experimental techniques such as terahertz (THz) spectroscopy, which monitors various transport mechanisms with femtosecond precision, serve as an ideal probe for unveiling the potentially divergent dynamical behaviors of spin and orbital angular momentum.

## c. Terahertz time domain spectroscopy (THz-TDS)

Terahertz (THz) radiation, spanning 0.1 to 30 THz, offers significant potential for fundamental research and technological innovation[8–16]. Its low energy aligns with key material excitations, including electron transport[17], plasmon excitation[18,19], Cooper pairs[13], phonons[11,20,21], and magnons[22–24]. The ultrafast timescale of THz pulses enables real-time probing of these dynamics, facilitating the study of ultrafast processes and decoupling angular momenta in solids[3,25]. THz radiation provides powerful tools for exploring charge carrier behavior, phase transitions, and collective phenomena, with applications ranging from ultrafast electronics to quantum computing and high-speed communications.

## d. Generation of THz radiation: Ultrafast spintronics and femtomagnetism

The recent development in ultrafast spintronics[26] and femtomagnetism[27] paved the way towards a promising Terahertz emitter[12,26–31]. It consists of a heterostructure of ferromagnetic and nonmagnetic material. Upon ultrafast excitation of the heterostructure, an ultrafast spin current is generated[27–29] which superdiffuses[29,32] into the adjacent nonmagnetic layer. If the adjacent nonmagnetic layer is a heavy metal like Pt or W with high spin orbit coupling, the out of plane spin current converts into in-plane charge current through inverse spin hall effect (ISHE)[12,28]. The resultant transient charge current emit electromagnetic radiation in THz frequencies[8,12,14–16,26,28,29,33–35]. These THz emitters utilize the spin angular momentum (SAM) of electrons to transport information through spin currents, which are then converted into detectable charge currents. However, in recent years, orbital angular momentum (OAM) has garnered significant interest in orbitronics research[1,2,5,36–41]. A key advantage of OAM over SAM is that it can assume arbitrarily larger values[4,37] in all the transition metals. Moreover, unlike spin-to-charge conversion, which relies on SOC, orbital-to-charge conversion can occur even in materials lacking SOC, including widely available light metals. Consequently, OAM does not experience SOC-induced scattering, potentially allowing it to propagate over longer distances compared to SAM, although the true extent of this advantage remains under active debate (see Section 2b). Thus, the utilization of reliable orbital current sources such as Ni and Co is essential for the development of orbitronic THz devices, offering a promising pathway toward next-generation electronic and optoelectronic technologies.

Several recent reviews and perspectives have addressed orbital Hall effects and orbital torques from a static or dc perspective[42–44]. In contrast, this Review focuses specifically on the ultrafast and dynamical aspects of orbitronics. We review the emergence of THz optorbitronics, where terahertz emission spectroscopy is used to disentangle spin and orbital transport on femtosecond timescales, quantify orbital diffusion lengths, and demonstrate active optical control of orbital currents. Beyond present demonstrations in metallic heterostructures, we critically examine emerging opportunities in three interesting directions:

1. Direct orbital current sources in graphene based orbital ferromagnets and altermagnets.
2. Potential of electrical tunability orbital transport.

3. The engineering of interfaces to maximize orbital to charge conversion efficiency. These future directions position THz optorbitronics as a powerful tool for both fundamental studies of orbital lattice coupling and the development of next generation ultrafast orbitronic devices.

## 2. Orbitronic THz emitter

### a. Working principle

As illustrated in Figure 2(a-e), when an FM/NM heterostructure where the ferromagnetic (FM) layer exhibits strong spin-orbit coupling (SOC), is excited by an ultrafast laser pulse, the initially generated spin current is converted into an orbital current due to the intrinsic $\langle L.S \rangle$ correlation within the ferromagnet. This process results in the injection of both spin current $(J_S)$ and orbital current $(J_L)$ into the adjacent nonmagnetic metal (NM) layer. Through the mechanisms of orbital-to-charge (LCC) and spin-to-charge (SCC) conversion, a transient charge current $(J_C(t))$ is generated, serving as the primary source of THz radiation. The resulting THz electric field is directly proportional to the transient sheet charge current $(J_C(t))$, as described by the following equation[25]

$$E(t) \propto J_C(t) \propto \int_{-d_{FM}}^{d_{NM}} dz [\theta_{LC}(z) J_L(z,t) + \theta_{SC}(z) J_s(z,t)] \qquad (1)$$

Here $d_{FM}$ and $d_{NM}$ denote the thickness of the ferromagnet and nonmagnetic material layer respectively and $\theta_{LC}$ and $\theta_{SC}$ denote the orbital and spin hall angle, which dictate the efficiency of the orbital to charge (LCC) and spin to charge conversion (SCC). This proportionality holds in the thin-film limit ( $d_{FM} + d_{NM} \ll c/\omega_{THz}$ ) and neglects impedance mismatch effects, which can modify the absolute THz amplitude but not the relative spin and orbital contributions. The equation excludes the minor process of THz emission such as ultrafast demagnetization and anomalous hall effect due to the ferromagnet, as these contributions can be experimentally disentangled from the primary photocurrent mechanism. Several research groups have shown orbital transport-based THz emission from magnetic heterostructure involving strong SOC nonmagnetic metal like Ni/W/CuO[25] as shown in Figure 2(a), Ni/Pt & Ni/Ru[3] as shown in Figure 2(b), and weak SOC based nonmagnetic metal like Ni/Cu[45], Co/Ti & Co/Mn[46] (Figure 2(c-d)) and NiFe/Nb[47] (Figure 2(e)). More recently, Guo et al.[48] reported THz emission from Ni/Ti bilayers driven by the inverse orbital Hall effect, extending the orbitronic THz emitter concept to the light 3d metal Ti where SOC is negligible. Shahriar and Elezzabi demonstrated[49] THz generation from Mo-based nanolayers via both the inverse orbital Hall and orbital Rashba–Edelstein effects, further broadening the accessible material palette. Additionally, Zhou et al.[50] have observed long-range orbital transport in Ru using Co/Ru and Co/Pt/Ru heterostructures, in which the THz signal persists even for ultrathick Ru layers and exhibits spectral broadening and delayed evolution incompatible with conventional spin transport. Strikingly, Zhou et al.[51] have also demonstrated IOHE-driven THz emission from Fe/Pt/W trilayers, showing that even a ferromagnet with quenched orbital moment (Fe) can serve as an orbital source when an intermediate high-SOC layer (Pt) mediates spin-to-orbital conversion. These results collectively demonstrate that orbitronic THz emission is not limited to a narrow class of Ni- or Co-based systems but extends to a broad range of FM/NM combinations spanning 3d, 4d, and 5d metals. The linear delay in the arrival of THz pulses and change in pulsewidth of the emitted THz pulse with increase in thickness of the nonmagnetic layer in case of Ni/W[25] (Figure 2(a)), Ni/Pt[3] (Figure 2(b)) and Ni/Cu[45] (Figure 2(c)) is a clear signature of long range ballistic transport of orbital current. Seifert et al.[25] modelled the ballistic long-range transport of the OAM in the NM layer by assuming a $\delta(t)$ like transient OAM accumulation in FM layer generating an electron wave packet. The orbital current accumulation and the orbital current density can be written as[25]

$$\mu_L(z,t) \propto \frac{1}{\sqrt{Dt}} exp\left(-\frac{z^2}{4Dt}\right) \tag{2}$$

$$r_z(t) \propto \Theta(t)\frac{z}{t}\frac{1}{\sqrt{Dt}} exp\left(-\frac{z^2}{4Dt}\right) \tag{3}$$

Here $\mu_L(z,t)$ is the orbital current accumulation in the FM layer

$r_z(t)$ = Orbital current density

$D$ = Diffusion coefficient which governs nature of the transport

$\Theta$ is the Heaviside step function

$z$ = thickness of the total layer where $z = 0$ is the FM/NM interface

Like the ferromagnet[48,49], antiferromagnetic insulators like $\alpha - Fe_2O_3$ can be a good source of orbital current through magnon transport as shown in Figure 3(a)[52]. In these systems, due to ultrafast photoexcitation, magnons are pumped to the adjacent high SOC nonmagnetic metal and are converted to coupled spin orbit current. These currents propagate toward the interface and IOREE at the interface generates charge current orthogonal to the spin-orbit coupled current. A similar approach can also be followed in case of a ferromagnet nonmagnetic heterostructures to enhance the effect of orbital transport as shown in Figure 3(b) where an intermediate high SOC layer could be used to convert the transported spin current from ferromagnet into orbital current. The orbital current can now travel to adjacent light metal and convert to charge current to generate THz. As a result, an enhancement in THz is observed when a light metal is introduced adjacent to a Co/W heterostructure as shown in Figure 3(b).

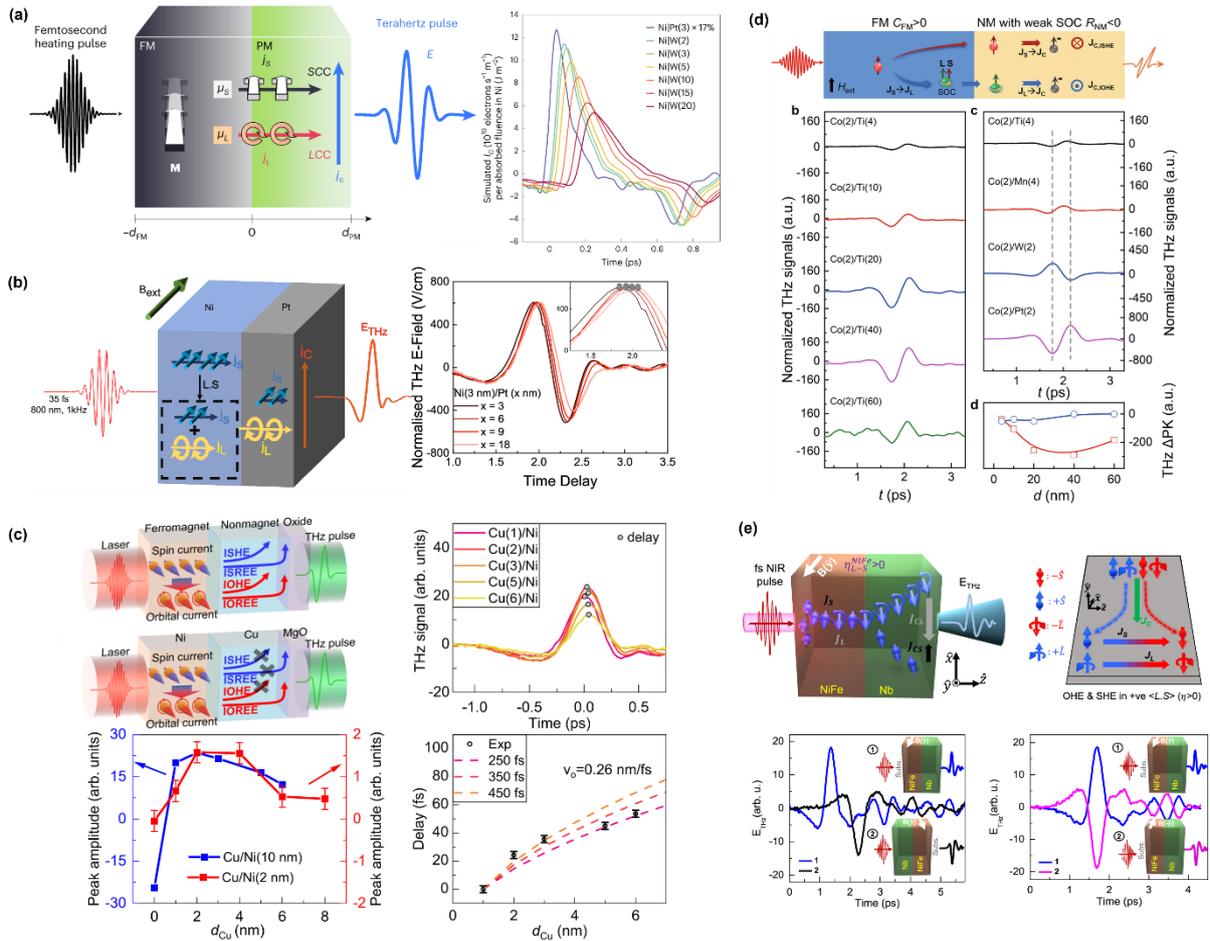

**Figure 2: THz emission from FM/NM Orbitronic heterostructure** (a) From Ni/W/CuO heterostructure through IOREE[25] (b) Ni/Pt heterostructure through IOREE[3] (c) From Ni/Cu/MgO heterostructure through IOREE[45] (d) From Co/Ti and Co/Mn heterostructure through IOHE[46] (e) NiFe/Nb heterostructure through IOHE[47]

## b. Long range ballistic orbital transport: A debate

In typical spintronic THz emitter, the length scale of the spin transport in nonmagnetic metal is <10 nm, due to the high SOC. A central question in orbitronics is the characteristic length scale over which OAM can be transported in metallic systems. Two strikingly different experimental answers have emerged. It has been shown that the orbital current can travel upto tens of nm in W[25], Ta[46] and Pt[3]. The long-range transport can be shown using the delay in the arrival of THz pulse as the thickness of the NM is increased as shown in Figure 4(a), 4(b) and 4(c). Due to long range transport of orbital current, a larger angular dispersion is expected as shown in Figure 4(a), resulting in an increase in pulse width of emitted THz pulse. The thickness of NM vs delay (Figure 4(b) and 4(c)) and thickness of NM vs pulse broadening (Figure 4(d) and 4(e)) in case of Ni/W (x nm) and Ni/Pt (x nm) is shown in Figure 4(b)-(e). The linear variation of the delay indicates the ballistic nature of orbital transport at least till x = 20 nm in W and x = 18 nm in Pt experimentally. The slope of the linear shift vs thickness plot gives the indication of orbital current velocity which is calculated to be around 0.25 nm/fs in case Ni/W[25], 0.18 nm/fs in Ni/Pt[3] and 0.26 nm/fs in Ni/Cu[45].

However, recent work by Guan et al.[53] fundamentally overturns this paradigm through the use of high-resolution terahertz emission spectroscopy combined with wedge shaped heterostructures that enable sub nanometer thickness control. Their systematic investigation demonstrates that the effective orbital diffusion length is dramatically shorter than previously assumed, measuring approximately 0.36 nm in W and 0.94 nm in Pt. These values imply localization on the scale of one to a few unit cells rather than transport over tens of nanometers.

At present, there is no consensus on the true magnitude of orbital transport lengths in these systems. Both the long-range values (~18–80 nm) and the ultra-short-range values (<1 nm) reported are based on careful experiments, yet they differ by one to two orders of magnitude. This fundamental uncertainty impacts interpretation of orbital transport efficiency and design principles for orbitronic devices and necessitates a substantial reorientation of theoretical and experimental orbitronics strategies. Experimental techniques like temperature dependent studies, complementary spatial probes as well as interfacial engineering could help resolve this controversy.

**Theoretical context for two competing paradigms**: The experimental discrepancy described above is mirrored by a fundamental theoretical debate that has recently crystallized around two opposing paradigms, as outlined in a recent perspective by Liao and Otani[54]. The first, the "hot spot" model proposed by Go *et al*[41]., suggests that genuine long-range OAM transport arises from symmetry-protected near-degenerate orbital states at specific points or lines in momentum space. At these hot spots, coherent superpositions of orbital states such as $(|d_{xy}\rangle + i|d_{yz}\rangle)/\sqrt{2}$ are resistant to the relaxation mechanisms that ordinarily quench OAM, enabling electrons near these k-space regions to carry orbital angular momentum over mesoscopic distances. Within this framework, both diffusive and ballistic transport regimes are accommodated. The former involves multiple scattering events with OAM retention during the orbital lifetime, while the latter produces the linear propagation delays observed in THz emission experiments. A key limitation of this model, however, is that hot spots occupy a small fraction of the Brillouin zone volume, raising the question of whether they can dominate transport without additional coherence mechanisms such as band-structure matching or quantum tunnelling. Furthermore, Tang and Bauer[55] have shown that the intrinsic OHE can be fully suppressed by arbitrarily weak disorder within the Born approximation, casting doubt on whether hot-spot-mediated transport survives in realistic polycrystalline samples. However, whether this result extends to realistic disorder strengths and to the extrinsic OHE regime remains an open question.

The second paradigm, the "local generation" model, takes the more radical position that genuine long-range OAM transport is not possible. This view is supported by first-principles scattering calculations by Rang and Kelly[56,57], who demonstrated that injected orbital currents decay within a few atomic layers (<1 nm) in disordered transition metals, consistent with conventional orbital quenching. Within a quantum kinetic theory developed by Valet et al.[58,59], the total OAM density is decomposed into an intraband component (the true orbital current, which is indeed short ranged) and an interband component (a local quantum coherence between electronic bands). In this framework, the experimentally observed long-range OAM accumulation does not represent transported orbital current but is instead locally generated wherever spatial gradients of the charge current exist. Near surfaces and interfaces, the charge current distribution is inherently non-uniform due to boundary conditions, giving rise to a finite "electron flow vorticity" $\nabla \times \mathbf{j}_{charge}$ that acts as a local source of orbital polarization. The apparent long characteristic decay length then reflects the spatial extent of charge current gradients. This charge current is typically governed by the electron mean free path rather than an intrinsic OAM diffusion length. Notably, recent work also suggests that grain boundaries and defects can locally generate OAM via the orbital Rashba-Edelstein effect[60], effectively reducing the distance over which OAM needs to propagate and further enhancing the apparent orbital transport length.

The implications of these two paradigms for orbitronic device design are fundamentally different. If the hot-spot picture is correct, material optimization should focus on engineering band structures with robust near-degenerate orbital states at the Fermi surface, thus favouring single-crystalline, high-purity materials with specific crystal symmetries. If local generation dominates, the emphasis shifts to geometric and interfacial engineering to maximize charge current gradients and electron flow vorticity, and the choice of material would be dictated primarily by the electron mean free path rather than orbital band structure features. It is also possible that both mechanisms coexist, with their relative importance dictated by material properties and experimental conditions. Several experimental strategies have been proposed to distinguish these paradigms: (i) systematic transport measurements on high-quality single crystals with controlled crystallographic orientations, where the hot-spot model predicts strong anisotropy while local generation does not; (ii) time-domain THz experiments testing whether local generation models can reproduce the linear propagation delays observed by Seifert *et al.* and Mishra *et al.*, given that the establishment of current gradients is limited by the momentum relaxation time; and (iii) nanostructures incorporating geometric constrictions or bends to deliberately engineer current vorticity, where geometry-dependent OAM signals independent of a distant source would support the local generation mechanism.

**Implications for the field:** Until the transport length scales are definitively established, quantitative predictions for orbital transport efficiency in practical devices remain uncertain. The long-range transport picture implies that thick nonmagnetic spacers could be used to engineer orbital pulses and directionality, whereas the ultra-short-range picture implies that orbital transport is highly sensitive to interfacial quality and limited to nanometer distances. Both scenarios have important implications for device engineering, but they cannot be simultaneously true. Resolution of this discrepancy should therefore be a priority for the field.

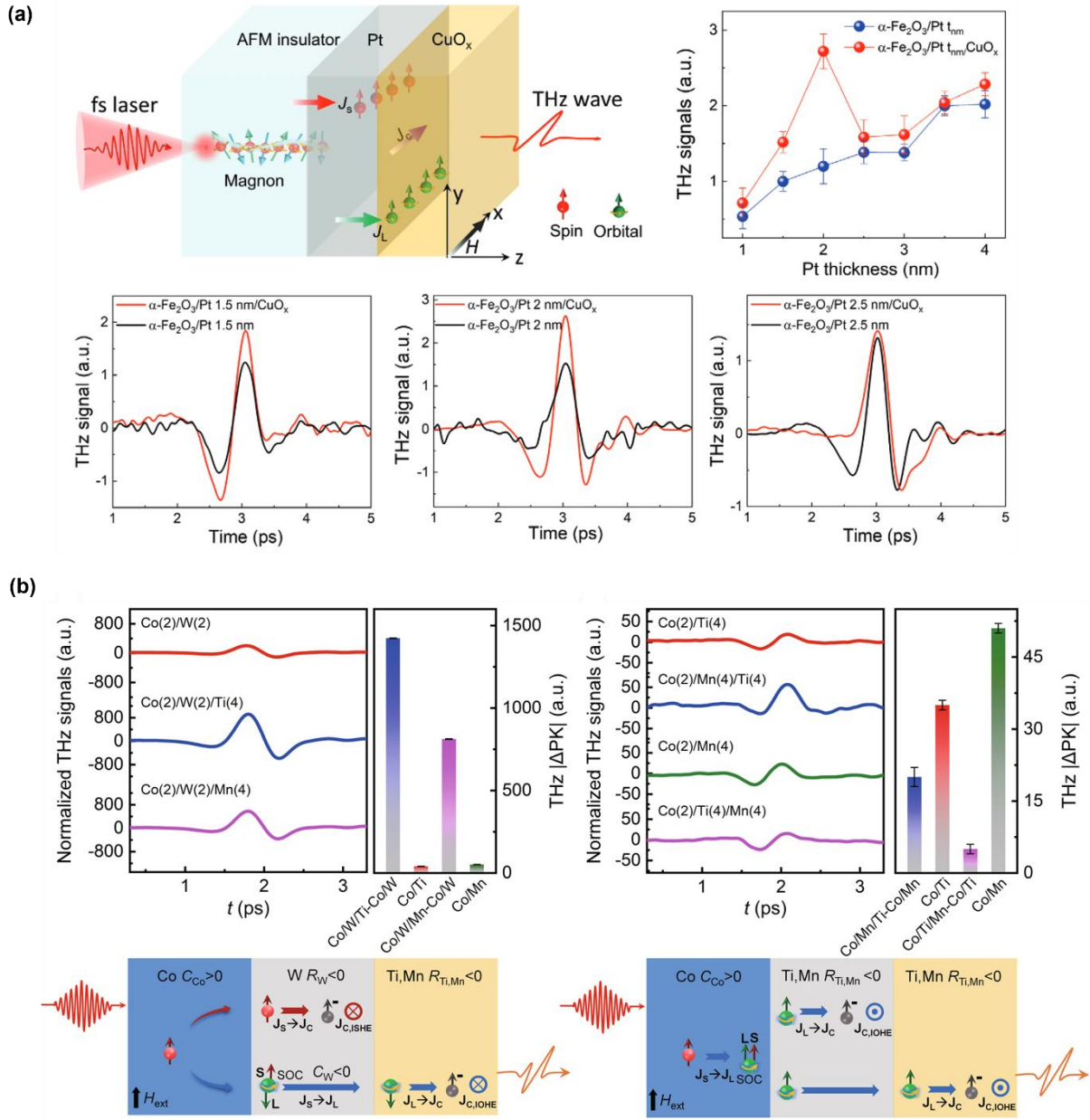

**Figure 3: Enhancement of THz emission using an intermediate high SOC layer** (a) THz orbitronics emitter using an antiferromagnetic insulator as an orbital current source. Here Pt is layer is used as a magnon to coupled spin-orbital current converter and Pt/ $CuO_x$ interface as an orbital to charge converter; (b) Enhancement of the THz emission due to spin to orbit conversion in a high SOC intermediate layer. Here W layer is used as a spin to orbital converter to enhance the THz emission.

## c. Optorbitronics: Optical control of orbital transport

Unlike spin angular momentum, orbital angular momentum couple to the lattice very efficiently through $\boldsymbol{L} - \boldsymbol{K}$ coupling[3,61]. The orbital angular momentum of the non-localized electrons interacts with the lattice through crystal field potential with following continuity equation

$$\frac{\partial}{\partial t}\langle L \rangle = \langle F^L \rangle + \frac{1}{i\hbar}\langle [L, V_{CF}] \rangle + \langle \lambda S \times L \rangle \qquad (4)$$

Here $F^L$ is the orbital flux term, $\langle [L, V_{CF}] \rangle$ describes the transfer of angular momentum between orbital and crystal through crystal field potential $V_{CF}$ and $\lambda S \times L$ describes the mutual transfer between spin and orbital angular momentum through spin orbit coupling. Upon photoexcitation of the orbitronic emitter with an ultrafast laser pulse, the electric field of the pulse perturbs the orbital wave function, driving it into a nonequilibrium state. As the laser fluence is increased, the magnitude of this perturbation grows, facilitating the extraction of angular momentum from the lattice and thereby modifying the orbital-crystal-field interaction. As a result, the velocity of orbital transport is enhanced.

Figure 4(f)-(l) offers a comprehensive illustration of the optical control of observed orbital transport within Pt[3] and Zr[62] layer with Ni and Co as the ferromagnet. Figures 4(f) and 4(g) show the shift in the emitted THz from Ni/Pt (3 nm) and Ni/Pt (6 nm) respectively with increase in fluence. The shift in the peak is shown in Figure 4(i) indicating a clear manipulation of orbital current velocity. With application of fluence, an initial right shift of the emitted pulse was observed. However, beyond a certain fluence threshold referred to as the critical fluence, there is an increase in orbital current velocity. Consequently, beyond this critical fluence, there is left shift in the arrival time of the THz pulse. Figure 4(i) and Figure 4(k) demonstrate that critical fluence is contingent on the thickness of the nonmagnetic Pt layer. As the thickness is increased, the carriers require more energy to overcome collisions to move ballistically over a larger distance, resulting in a higher critical fluence. The slope of the thickness versus the critical energy was calculated and termed as critical energy density and was calculated to be 343.7 $J/cm^2$. Similar behavior was also observed when the peak-to-peak time was recorded as shown in Fig 4(j). The change in orbital velocity is shown in Figure 4(h) where the orbital current velocity could be increased from 0.14 nm/fs to 0.18 nm/fs when fluence is increased from 191 $\mu J/cm^2$ to 1270 $\mu J/cm^2$.

A similar control of orbital current velocity was also shown in Co/Zr and Co/W/Cr heterostructure[62] as shown in Figure 4(l) and Figure 4(m). Figure 4(m) elucidates the critical fluence density for C/W/Zr and Co/W heterostructures to be 624.2 $J/cm^2$ and 440.6 $J/cm^2$ respectively.

Figure 4(n) demonstrates the mechanism of the orbital transport mechanism before and after reaching the critical fluence. Nonlocalized electrons, bearing information about orbital angular momentum, traverse through orbital hopping between nuclei within a solid. An increase in fluence increases the number of charge carriers thereby enhancing the number of collisions that impede the transport process. Nevertheless, beyond the critical fluence, nonlocalized electrons perturbed by laser fluence absorb additional angular momentum from the lattice according to equation 4[61,63,64], as illustrated in Figure 4(n-ii). As we apply laser fluence, the local magnetic moment couples with the spin of the nonlocalized conduction electrons through exchange interaction[15,64], thus creating a spin current. Due to high spin orbit correlation of Ni near Fermi level, there is angular momentum transfer between spin and orbital which can be explained by the cross product of S and L ($\langle \lambda S \times L \rangle$) as shown in equation 4. Additionally, due to the electric field of the applied laser pulse, a perturbation in the orbital wave function is induced, thus creating a non-equilibrium state. Consequently, the non-equilibrium orbital wave function takes angular momentum from the lattice through the crystal field potential $V_{\mathrm{CF}}$ explained by $\frac{1}{i\hbar}\langle [L, V_{CF}] \rangle$ in the equation 4, enabling the orbital current to surpass collisions, facilitating a more rapid transport.

While the **L–K** coupling mechanism provides a consistent interpretation of the data, alternative explanations cannot be fully excluded at present. At fluences exceeding 1000 $\mu J/cm^2$, significant electron and lattice heating occurs, which could modify the superdiffusive transport characteristics or induce Fermi-surface thermal smearing that alters the relative orbital and spin contributions. Interface degradation or interdiffusion at elevated fluences might also produce apparent timing shifts. The observation that no fluence-dependent shift is seen in spin-dominated NiFe/Pt (Figure 4(i)) provides an important control, arguing against purely thermal or spin-mediated origins for the effect observed in Ni/Pt. Nevertheless, distinguishing these scenarios conclusively will require fluence-dependent measurements at cryogenic temperatures, where lattice heating is suppressed, as well as post-measurement structural characterization of the interfaces.

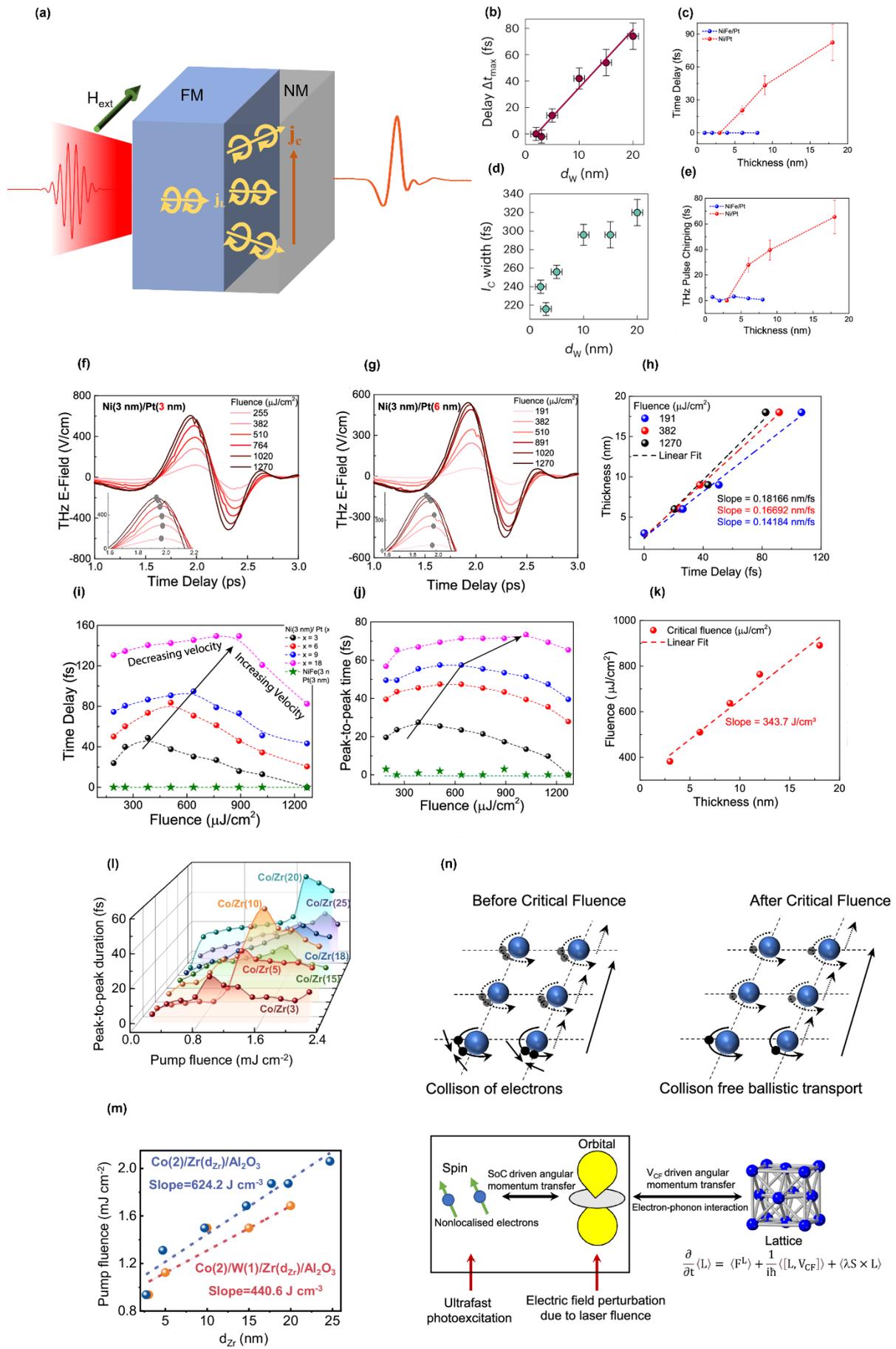

**Figure 4: Long range active ballistic orbital transport**: (a) Mechanism of long range ballistic transport in FM/NM heterostructure; Linear variation of delay vs thickness in (b) Ni/W (c) Ni/Pt; Pulse broadening with increase in thickness of NM layer in (d) Ni/W, (e) Ni/Pt; THz emission from Ni (3 nm)/Pt (x nm) at different fluence when (f) x = 3, (g) x = 6, (h) Extracted time delay with thickness at different fluences. The slope of the linear fitted line can be termed as orbital current velocity which could be tuned from 0.14 nm/fs to 0.18 nm/fs as the fluence was increased from 191 $\mu J/cm^2$ to 1270 $\mu J/cm^2$ (i) Extracted time delay for different fluence for Ni (3 nm)/Pt (x nm) with x = 3, 6, 9, 18. Initially, the shift is towards right indicating the decrease in orbital velocity and after a critical fluence, delay starts decreasing with increase in fluence showing swifter orbital transport; Similar shift is not seen in spin transport as shown in NiFe (3 nm)/Pt (3 nm). (j) THz pulse chirping, for different fluence is shown for Ni (3 nm)/Pt (x nm) with x = 3, 6, 9, and 18. A thickness-dependent critical fluence similar to (i) can be seen. (k) Extracted critical fluence with variation in thickness. The slope of the fitted straight line indicates the "critical energy density". (l) THz pulse chirping, for different fluence is shown for Co/Zr/Al$_2$O$_3$. A thickness-dependent critical fluence similar to (j) can be seen (m) "critical energy density" for Co/Zr/Al$_2$O$_3$ and Co/W/Zr/Al$_2$O$_3$

## 3. Future potential and discussion

### a. Disentanglement of spin and orbital diffusion length

In most metals, the orbital diffusion length ($\lambda_L$) exceeds its spin counterpart ($\lambda_S$). However, direct experimental quantification of ($\lambda_L$) and ($\lambda_S$) remains challenging due to the lack of techniques with sufficient temporal resolution and selectivity to the different angular momentum channels. The femtosecond time resolution of THz spectroscopy gives us the opportunity to exploit the distinct behavior of spin and orbital angular momentum on the ultrafast timescale.

Figure 5(a–b) shows the comparison of THz transmission and THz emission spectroscopy of Ni/Pt and NiFe/Pt heterostructures for different thicknesses of Pt. At thicknesses exceeding the relaxation lengths of all contributing transport entities, both THz transmission and emission are expected to exhibit similar trends, reflecting saturation of the underlying transport processes. In the case of Ni/Pt (x nm), the emitted THz pulse shows a linear decrease from x = 3 nm to 18 nm, indicating dispersion and attenuation, in contrast to the exponentially decreasing transmitted peak-to-peak THz pulse in the same thickness range as shown in Figure 5(a). This suggests a relaxation length exceeding 18 nm for the dominant orbital contribution to the emitted signal. We assume that beyond the relaxation length, the emitted signal is expected to saturate, transitioning to an exponential behavior analogous to transmission.

Conversely, in NiFe/Pt, where orbital contributions are suppressed and spin transport dominates, both THz emission and transmission exhibit a consistent exponential decay beyond the spin relaxation length (~1.2 nm) (Figure 5(b)). The minor reduction in the emission decay rate may arise from secondary spin current generation mechanisms[35]. These observations highlight the capability of THz spectroscopy to distinguish between spin and orbital transport dynamics and establish a framework for estimating their respective diffusion lengths in high-SOC systems.

**Methodological considerations**: The comparison between Ni/Pt and NiFe/Pt implicitly assumes that Ni is orbital-dominant whereas NiFe is spin-dominant and that the only major difference between the two systems is the relative weight of spin and orbital channels. In practice, Fe substitution also modifies the magnetic damping, the SOC strength and potentially the interface morphology, all of which can affect the THz response. A more rigorous separation of spin and orbital diffusion lengths will benefit from temperature-dependent measurements, in which spin- and orbital-mediated contributions are expected to show different temperature scaling, and from complementary probes such as time-resolved magneto-optical Kerr measurements or transport-based detection of orbital accumulation.

A similar approach was followed in Kumar et al[47]. to calculate the spin and orbital diffusion length in the light metal Nb. Additionally, they performed a temperature-dependent study of the THz emission to disentangle the spin and orbital contributions to the THz signal. Such combined thickness- and temperature-dependent THz emission and transmission measurements, supported by appropriate transport modelling (e.g., superdiffusive or Boltzmann approaches), provide a powerful route to extract $(\lambda_L)$ and $(\lambda_S)$ across a wide range of materials.

Based on these developments, we propose a standardized experimental protocol for disentangling spin and orbital diffusion lengths in a given NM material: (i) thickness-dependent THz emission and transmission measurements to identify the functional form (linear versus exponential) of the signal decay; (ii) temperature-dependent THz emission to exploit the different temperature scaling of spin-mediated (magnon- and phonon-dependent) and orbital-mediated (crystal-field-dependent, largely temperature-independent) contributions; (iii) control measurements using NiFe or Fe-based ferromagnets to benchmark the spin-only response; and (iv) fluence-dependent studies to probe the orbital–lattice coupling regime and confirm the orbital origin of any anomalous transport signatures

Looking ahead, applying this protocol to different 3d, 4d, 5d, and 4f systems, as well as altermagnets and 2D materials, will be crucial for establishing systematic trends in orbital vs. spin diffusion lengths and identifying materials with exceptionally long-range orbital transport suitable for THz orbitronic devices.

**b. Direct source of ultrafast orbital current: orbital ferromagnetism in graphene, altermagnets and its ultrafast demagnetization**

The existence of orbital ferromagnetic states in engineered graphene systems has been experimentally established in several material platforms. Among these, twisted bilayer graphene with a twist angle close to 1.68° has emerged as a particularly promising system. In this configuration, researchers have reported the observation of significant orbital magnetic moments, on the order of 10.7 µB per moiré supercell[65,66]. These moments originate from chiral circulating current loops that are confined within the moiré potential landscape, leading to spontaneous time-reversal symmetry breaking and valley-selective magnetic ordering. This form of magnetism is fundamentally distinct from conventional spin-based ferromagnetism, as it arises purely from the motion of charge carriers and their associated orbital degrees of freedom, without relying on spin polarization.

Theoretical studies further predict that rhombohedral multilayer graphene may host even stronger orbital magnetism due to its flat electronic bands and stacking-order-enhanced interactions. The tunability of such systems through external electric fields, gating, or doping offers a powerful mechanism for dynamic control of orbital magnetism. These findings firmly establish graphene-based orbital ferromagnets as a novel platform hosting sizeable static orbital magnetization. However, ultrafast orbital-current generation and THz emission from these systems have not yet been experimentally demonstrated. Realizing gate-tunable THz emitters based on orbital ferromagnetic graphene will require future experiments that combine ultrafast optical excitation with sensitive THz detection and careful control of disorder and device geometry. Nevertheless, the combination of strong orbital order and electrical tunability makes these systems highly promising emerging candidates for reconfigurable orbitronic THz devices.

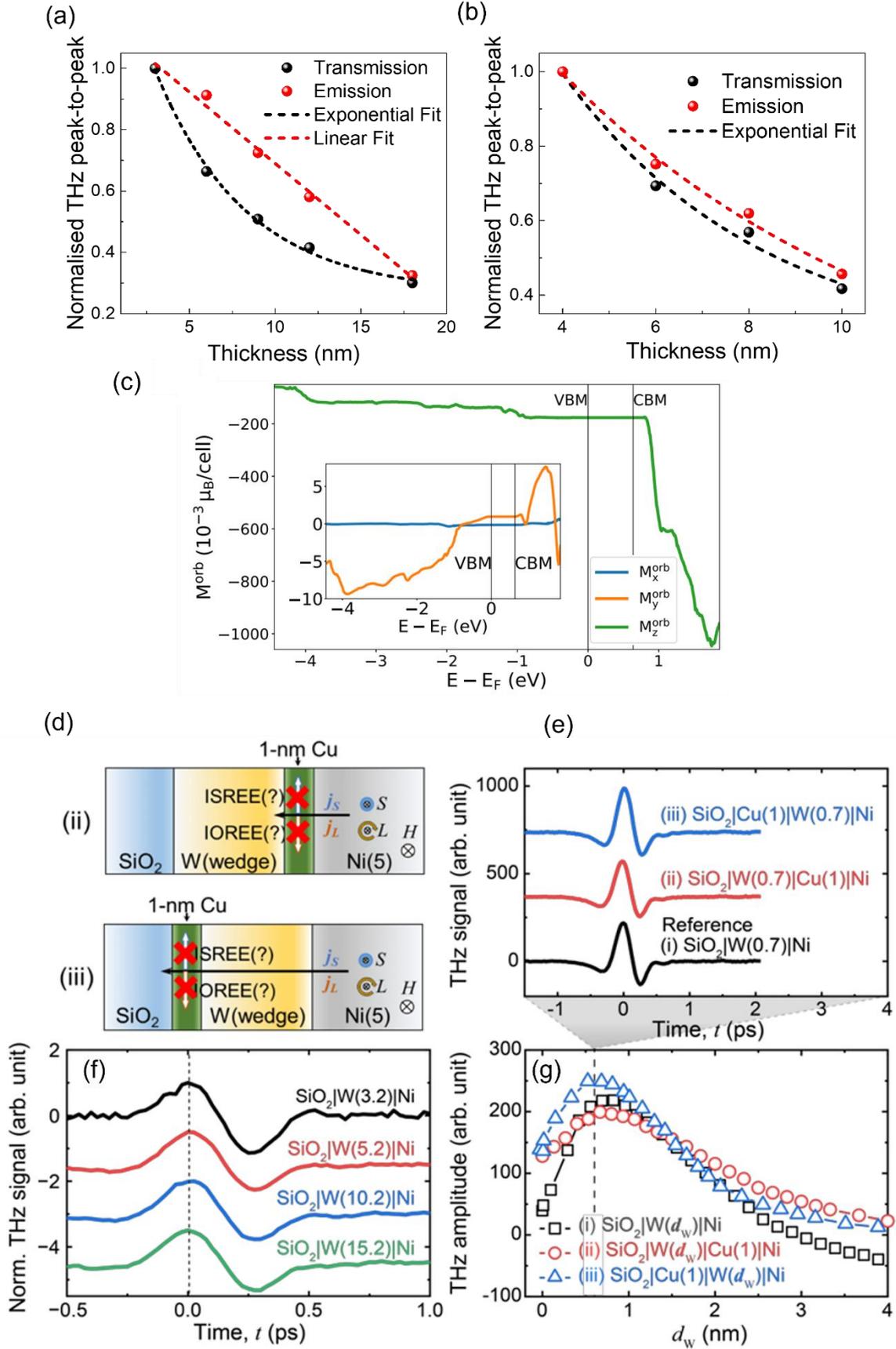

**Figure 5: Future potential of THz optorbitronics:** (a) Ni (3 nm) / Pt (x nm) showing a linear decay in the emitted THz pulse in contrast to an exponential decay in the 1 transmitted THz pulse indicating a longer transport phenomenon; (b) NiFe (3 nm)/ Pt (x nm) showing an exponential decay in both emitted

and transmitted THz pulse indicating a shorter transport phenomenon; (c) Orbital magnetization in MnTe; (d) Schematic illustrations of configurations with 1-nm Cu spacer layers inserted at different interfaces: configuration (ii) with Cu at the W|Ni interface, and configuration (iii) with Cu at the SiO2|W interface. Cu insertion is designed to disrupt interfacial conversion processes such as the ISREE and IOREE; (e) Experimental terahertz waveforms from three sample configurations: (i) SiO2|W(0.7)|Ni, (ii) SiO2|W(0.7)|Cu(1)|Ni, and (iii) SiO2|Cu(1)|W(0.7)|Ni; (f) Terahertz amplitudes extracted at t=0 as a function of dW from all three configurations, showing consistent peak behavior near dW ≈ 0.7-0.8 nm; (g) Normalized terahertz waveforms from SiO2|W|Ni heterostructures with varying dW. Peak positions are denoted by a dashed line.

In parallel, recent theoretical and experimental work has shown that certain altermagnets, such as g-wave MnTe,[67,68] can host dominant orbital magnetization and exhibit strong coupling between orbital degrees of freedom, crystal symmetry, and strain. Strain-controlled phase transitions between antiferromagnetic and altermagnetic states have been predicted and, in some cases, demonstrated, providing an additional knob to manipulate orbital ferromagnetism. These altermagnetic materials offer a complementary pathway to generate ultrafast orbital currents via the demagnetization of orbital order and the dynamics of altermagnons.

At present, no orbital magnetization has been directly measured in altermagnets and THz emission from altermagnetic orbital sources has not been reported. Demonstrating ultrafast orbital-current generation in these materials should be viewed as a medium-term goal requiring advances in crystal growth, strain control, and interface engineering.

The combination of graphene-based orbital ferromagnets and altermagnets thus provides a rich materials platform for direct ultrafast orbital-current generation. Future THz orbitronic experiments could, for example, probe gate-tunable THz emission from orbital ferromagnetic graphene, or strain- and temperature-dependent THz emission from altermagnets, thereby directly accessing orbital demagnetization and altermagnon dynamics on femtosecond timescales.

## c. Electrical tunability of orbital current in FM/NM heterostructures

Since spin and orbital angular momenta are coupled through spin-orbit interaction, experimental techniques developed to manipulate spin currents can, in principle, be extended to control orbital transport. Recent experimental progress has demonstrated the ability to electrically modulate spin currents in terahertz spintronic emitters through the use of ferroelectric substrates[8]. Building on this concept, similar approaches may be extended to enable the control of orbital angular momentum, offering a pathway to develop electrically tunable orbitronic devices.

Another route is to exploit the electrical control of orbital ferromagnetism that has already been reported in graphene-based systems, where gate voltages tune the occupation of flat bands and the strength of orbital order[65,66,69]. Moreover strain control[68,70,71] of magnetic phases in altermagnets have opened up avenues for electrical tunability of orbital magnetism in these materials. These findings lead to possibilities of electrically and mechanically tunable orbitronic THz emitters with enhanced functionalities. Combining optical, electrical, and strain control offers a multidimensional parameter space for tailoring orbital current amplitude, velocity, and polarization, enabling active THz modulators and reconfigurable THz sources based on orbital transport.

### d. Role of interface in orbital transport

The role of interfaces has been discussed in detail in the context of spin transport and spin-based THz emitters, where phenomena such as the interfacial Rashba–Edelstein effect, spin-memory loss, and interfacial skew scattering play a crucial role in determining the efficiency of spin injection and spin-to-charge conversion[30,72,73]. However, similarly comprehensive studies are not yet available for orbital-based THz emitters. Notably, since orbital current can travel longer distances in the nonmagnetic metal layer than spin transport, it is particularly important to understand and engineer the role of interfaces in orbital injection and orbital-to-charge conversion. Interfaces with strong inversion-symmetry breaking and significant orbital hybridization are expected to host the orbital Rashba effect and its inverse, enabling efficient orbital-to-charge conversion via the orbital inverse Rashba-Edelstein effect. Recent experiments have already reported direct observation of this effect at specific metal/oxide interfaces, highlighting the potential of interface engineering for orbitronic devices. In particular, Xu et al.[74] have demonstrated THz generation via the inverse orbital Rashba–Edelstein effect at the $Ni/CuO_x$ interface, showing that oxygen-induced orbital hybridization tunes the IOREE efficiency. This result provides the first example of chemically controllable orbital-to-charge conversion at an interface, establishing a concrete strategy for interface engineering in orbitronic THz emitters

Future work should therefore focus on systematically characterizing how interfacial chemistry, crystallographic orientation, roughness, and adjacent oxide or 2D layers influence orbital injection and conversion in FM/NM heterostructures as shown in Figure 5(d-g)[53]. Combining THz emission spectroscopy with complementary probes such as angle-resolved photoemission, magnetotransport, and *ab initio* calculations of orbital textures across interfaces will be essential. Such efforts will not only clarify the microscopic origin of orbital-to-charge conversion in existing orbitronic THz emitters but also guide the rational design of multilayer stacks that maximize orbital Hall currents, interface-mediated orbital-to-charge conversion, and overall THz emission efficiency[6,7,75].

| Platform | Key Properties | Demonstrated | Key challenges |
|---|---|---|---|
| **FM/NM heterostructures (Ni/W, Ni/Pt, Co/Ti, Ni/Cu)** | Orbital Current generation and long-range transport | Yes | Orbital current length scale debate |
| **Topological Dirac Materials** | Static orbital ferromagnetism | Yes (No THz studies) | Ultrafast orbital dynamics unknown; THz emission studies |
| **Altermagnets** | Orbital magnetism | Theoretical studies only | Crystal growth challenges; strain control; orbital magnetization not measured; THz sources not realized |
| **AFM insulators** | Magnon-to-orbital conversion | Magnon generation unknown | Orbital accumulation from magnons not demonstrated; efficiency unclear |
| **Electrically tunable sources** | Gate tunability of orbital response in graphene and altermagnets | Theoretical | Requires ultrafast gate switching; ultrafast THz generation not demonstrated |

**Table 1: Status of orbital current sources for THz optorbitronics**

## Acknowledgments

S.M. thanks Dr. Thomas Tan, Dr. James Lourembam, and Sambhu Jana for valuable discussions and suggestions.

## Author Contributions

SM, and R.S. conceived the review paper; S.M. and R.S. wrote the manuscript; R.S. lead the overall project.

## Competing Interests

The authors declare no competing interests

## References

1. Go, D., Jo, D., Kim, C. & Lee, H.-W. Intrinsic Spin and Orbital Hall Effects from Orbital Texture. *Phys. Rev. Lett.* **121**, 086602 (2018).
2. Go, D., Jo, D., Lee, H.-W., Kläui, M. & Mokrousov, Y. Orbitronics: Orbital currents in solids. *EPL* **135**, 37001 (2021).
3. Mishra, S. S., Lourembam, J., Lin, D. J. X. & Singh, R. Active ballistic orbital transport in Ni/Pt heterostructure. *Nat Commun* **15**, 4568 (2024).
4. Lee, D. *et al.* Orbital torque in magnetic bilayers. *Nat Commun* **12**, 6710 (2021).
5. Go, D. *et al.* Long-Range Orbital Magnetoelectric Torque in Ferromagnets. Preprint at http://arxiv.org/abs/2106.07928 (2022).
6. Park, J.-H., Kim, C. H., Rhim, J.-W. & Han, J. H. Orbital Rashba effect and its detection by circular dichroism angle-resolved photoemission spectroscopy. *Phys. Rev. B* **85**, 195401 (2012).
7. Park, S. R., Kim, C. H., Yu, J., Han, J. H. & Kim, C. Orbital-Angular-Momentum Based Origin of Rashba-Type Surface Band Splitting. *Phys. Rev. Lett.* **107**, 156803 (2011).
8. Agarwal, P., Huang, L., Ter Lim, S. & Singh, R. Electric-field control of nonlinear THz spintronic emitters. *Nat Commun* **13**, 4072 (2022).
9. Mishra, S. S., Tan, T. C., Gupta, M., Xiu, F. & Singh, R. Electrically tunable Floquet Weyl photon emission from Dirac semimetal $Cd_3As_2$. Preprint at https://doi.org/10.48550/ARXIV.2501.16498 (2025).
10. Yang, Y. *et al.* Terahertz topological photonics for on-chip communication. *Nat. Photonics* **14**, 446–451 (2020).
11. Zhang, B. *et al.* Electric-field Control of Giant Ferronics. Preprint at https://doi.org/10.48550/arXiv.2509.06057 (2025).
12. Seifert, T. *et al.* Efficient metallic spintronic emitters of ultrabroadband terahertz radiation. *Nature Photon* **10**, 483–488 (2016).
13. Srivastava, Y. K. *et al.* $YBa_2Cu_3O_7$ as a high-temperature superinductor. *Nat. Mater.* **24**, 883–890 (2025).
14. Agarwal, P. *et al.* Reconfigurable Chiral Spintronic THz Emitters. *Advanced Optical Materials* **12**, 2303128 (2024).
15. Agarwal, P. *et al.* Terahertz spintronic magnetometer (TSM). *Applied Physics Letters* **120**, 161104 (2022).
16. Jana, S., Mishra, S. S., Lourembam, J. & Singh, R. Ultrafast Control of Néel Vector in Collinear Antiferromagnet MnPt. *Advanced Science* **13**, e19395 (2026).
17. Dai, Z. *et al.* High Mobility 3D Dirac Semimetal ($Cd_3As_2$) for Ultrafast Photoactive Terahertz Photonics. *Adv. Funct. Mater.* **31**, 2011011 (2021).
18. Arezoomandan, S., Condori Quispe, H. O., Ramey, N., Nieves, C. A. & Sensale-Rodriguez, B. Graphene-based reconfigurable terahertz plasmonics and metamaterials. *Carbon* **112**, 177–184 (2017).
19. Chanana, A. *et al.* Manifestation of Kinetic Inductance in Terahertz Plasmon Resonances in Thin-Film $Cd_3As_2$. *ACS Nano* **13**, 4091–4100 (2019).
20. Dekorsy, T., Auer, H., Bakker, H. J., Roskos, H. G. & Kurz, H. THz electromagnetic emission by coherent infrared-active phonons. *Phys. Rev. B* **53**, 4005–4014 (1996).
21. Hou, L. *et al.* Temperature-dependent terahertz properties of carriers and phonons in the topological Dirac semimetal $Cd_3As_2$. *Phys. Rev. B* **108**, 115416 (2023).

22. Jost, D. *et al*. Chiral Altermagnon in MnTe. Preprint at https://doi.org/10.48550/arXiv.2501.17380 (2025).

23. Bose, A. *et al*. Tilted spin current generated by the collinear antiferromagnet ruthenium dioxide. *Nat Electron* **5**, 267–274 (2022).

24. Behovits, Y. *et al*. Terahertz Néel spin-orbit torques drive nonlinear magnon dynamics in antiferromagnetic Mn2Au. *Nat Commun* **14**, 6038 (2023).

25. Seifert, T. S. *et al*. Time-domain observation of ballistic orbital-angular-momentum currents with giant relaxation length in tungsten. *Nat. Nanotechnol.* https://doi.org/10.1038/s41565-023-01470-8 (2023) doi:10.1038/s41565-023-01470-8.

26. Walowski, J. & Münzenberg, M. Perspective: Ultrafast magnetism and THz spintronics. *Journal of Applied Physics* **120**, 140901 (2016).

27. Beaurepaire, E. Ultrafast Spin Dynamics in Ferromagnetic Nickel. *PHYSICAL REVIEW LETTERS* **76**, (1996).

28. Kampfrath, T. *et al*. Terahertz spin current pulses controlled by magnetic heterostructures. *Nature Nanotech* **8**, 256–260 (2013).

29. Seifert, T. S., Cheng, L., Wei, Z., Kampfrath, T. & Qi, J. Spintronic sources of ultrashort terahertz electromagnetic pulses. *Appl. Phys. Lett.* **120**, 180401 (2022).

30. Zhou, C. *et al*. Broadband Terahertz Generation via the Interface Inverse Rashba-Edelstein Effect. *Phys. Rev. Lett.* **121**, 086801 (2018).

31. Rouzegar, R. *et al*. Broadband Spintronic Terahertz Source with Peak Electric Fields Exceeding 1.5 MV/cm. *Phys. Rev. Applied* **19**, 034018 (2023).

32. Battiato, M., Carva, K. & Oppeneer, P. M. Superdiffusive Spin Transport as a Mechanism of Ultrafast Demagnetization. *Phys. Rev. Lett.* **105**, 027203 (2010).

33. Comstock, A. *et al*. Spintronic Terahertz Emission in Ultrawide Bandgap Semiconductor/Ferromagnet Heterostructures. *Advanced Optical Materials* **11**, 2201535 (2023).

34. Koleják, P. *et al*. 360° Polarization Control of Terahertz Spintronic Emitters Using Uniaxial FeCo/TbCo $_2$ /FeCo Trilayers. *ACS Photonics* **9**, 1274–1285 (2022).

35. Agarwal, P. *et al*. Secondary Spin Current Driven Efficient THz Spintronic Emitters. *Advanced Optical Materials* 2301027 (2023) doi:10.1002/adom.202301027.

36. Sala, G. & Gambardella, P. Giant orbital Hall effect and orbital-to-spin conversion in 3 d , 5 d , and 4 f metallic heterostructures. *Phys. Rev. Research* **4**, 033037 (2022).

37. Kontani, H., Tanaka, T., Hirashima, D. S., Yamada, K. & Inoue, J. Giant Orbital Hall Effect in Transition Metals: Origin of Large Spin and Anomalous Hall Effects. *Phys. Rev. Lett.* **102**, 016601 (2009).

38. Li, T. *et al*. Giant Orbital-to-Spin Conversion for Efficient Current-Induced Magnetization Switching of Ferrimagnetic Insulator. *Nano Lett.* acs.nanolett.3c02104 (2023) doi:10.1021/acs.nanolett.3c02104.

39. Jo, D., Go, D. & Lee, H.-W. Gigantic intrinsic orbital Hall effects in weakly spin-orbit coupled metals. *Phys. Rev. B* **98**, 214405 (2018).

40. Go, D. *et al*. Orbital Pumping by Magnetization Dynamics in Ferromagnets. Preprint at http://arxiv.org/abs/2309.14817 (2023).

41. Go, D. *et al*. Long-Range Orbital Torque by Momentum-Space Hotspots. *Phys. Rev. Lett.* **130**, 246701 (2023).

42. Fukami, S., Lee, K.-J. & Kläui, M. Challenges and opportunities in orbitronics. *Nat. Phys.* 1–7 (2025) doi:10.1038/s41567-025-03143-w.

43. Jo, D., Go, D., Choi, G.-M. & Lee, H.-W. Spintronics meets orbitronics: Emergence of orbital angular momentum in solids. *npj Spintronics* **2**, 19 (2024).

44. Wang, P. *et al*. Orbitronics: Mechanisms, Materials and Devices. *Advanced Electronic Materials* **11**, 2400554 (2025).

45. Xu, Y. *et al*. Orbitronics: light-induced orbital currents in Ni studied by terahertz emission experiments. *Nat Commun* **15**, 2043 (2024).

46. Wang, P. Inverse orbital Hall effect and orbitronic terahertz emission observed in the materials with weak spin-orbit coupling. *npj Quantum Materials* (2023).

47. Kumar, S. & Kumar, S. Ultrafast THz probing of nonlocal orbital current in transverse multilayer metallic heterostructures. *Nat Commun* **14**, 8185 (2023).

48. Guo, C. *et al*. Ferromagnetic $\mathrm{Ni}$ enhances terahertz emission through the inverse orbital Hall effect in $\mathrm{Ni}/\mathrm{Ti}$ bilayers. *Phys. Rev. Appl.* **24**, 024009 (2025).

49. Shahriar, B. Y. & Elezzabi, A. Y. Orbitronic terahertz emission from Mo-based nanolayers via the inverse orbital Hall and Rashba–Edelstein effects. *Appl. Phys. Lett.* **127**, 151103 (2025).

50. Chao, Z. *et al*. Long-range orbital transport and inverse orbital Hall effect in Co/Ru-based terahertz emitters. Preprint at https://doi.org/10.48550/arXiv.2602.04186 (2026).

51. Zhou, C. *et al*. Inverse orbital Hall effect induced terahertz emission enabled by a ferromagnet with quenched orbital moment in Fe/Pt/W trilayers. Preprint at https://doi.org/10.48550/arXiv.2602.08516 (2026).

52. Huang, L. *et al*. Orbital Current Pumping From Ultrafast Light-driven Antiferromagnetic Insulator. *Advanced Materials* **37**, 2402063 (2025).

53. Guan, T. *et al*. Evidence of Ultrashort Orbital Transport in Heavy Metals Revealed by Terahertz Emission Spectroscopy. Preprint at https://doi.org/10.48550/arXiv.2504.15553 (2025).

54. Liao, L. & Otani, Y. Long-range transport versus local generation in orbitronics. *Commun Phys* **9**, 48 (2026).

55. Tang, P. Role of Disorder in the Intrinsic Orbital Hall Effect. *Phys. Rev. Lett.* **133**, (2024).

56. Rang, M. & Kelly, P. J. Orbital Hall effect in transition metals from first-principles scattering calculations. *Phys. Rev. B* **111**, 125121 (2025).

57. Rang, M. & Kelly, P. J. Orbital relaxation length from first-principles scattering calculations. *Phys. Rev. B* **109**, 214427 (2024).

58. Valet, T., Jaffrès, H., Cros, V. & Raimondi, R. Quantum Kinetic Anatomy of Electron Angular Momenta Edge Accumulation. *Phys. Rev. Lett.* **135**, 256301 (2025).

59. Valet, T. Quantum kinetic theory of the linear response for weakly disordered multiband systems. *Phys. Rev. B* **111**, (2025).

60. Idrobo, J. C. *et al*. Direct observation of nanometer-scale orbital angular momentum accumulation. Preprint at https://doi.org/10.48550/arXiv.2403.09269 (2024).

61. Han, S. *et al*. Orbital Pumping Incorporating Both Orbital Angular Momentum and Position. *Phys. Rev. Lett.* **134**, 036305 (2025).

62. Xu, H. *et al*. Generation and manipulation of light-induced orbital transport in Co/Zr/Al2O3 heterostructure probed with ultrafast terahertz emission. *Commun Phys* **8**, 115 (2025).

63. Haney, P. M. & Stiles, M. D. Current-Induced Torques in the Presence of Spin-Orbit Coupling. *Phys. Rev. Lett.* **105**, 126602 (2010).

64. Go, D. *et al*. Theory of current-induced angular momentum transfer dynamics in spin-orbit coupled systems. *Phys. Rev. Res.* **2**, 033401 (2020).

65. Bhardwaj, V. *et al*. Gate-Tunable Orbital Magnetism and Competing Superconductivity in Twisted Trilayer Graphene Josephson Junctions. *ACS Appl. Mater. Interfaces* **17**, 69784–69794 (2025).

66. Sharpe, A. L. *et al*. Evidence of Orbital Ferromagnetism in Twisted Bilayer Graphene Aligned to Hexagonal Boron Nitride. *Nano Lett.* **21**, 4299–4304 (2021).

67. Chen Ye, C., Tenzin, K., Sławińska, J. & Autieri, C. Dominant orbital magnetization in the prototypical altermagnet MnTe. *Phys. Rev. B* **113**, 014413 (2026).

68. Karetta, B., Verbeek, X. H., Jaeschke-Ubiergo, R., Šmejkal, L. & Sinova, J. Strain-controlled g to d wave transition in altermagnetic CrSb. *Phys. Rev. B* **112**, 094454 (2025).

69. Imaging orbital ferromagnetism in a moiré Chern insulator. https://www.science.org/doi/10.1126/science.abd3190 doi:10.1126/science.abd3190.

70. Chakraborty, A., González Hernández, R., Šmejkal, L. & Sinova, J. Strain-induced phase transition from antiferromagnet to altermagnet. *Phys. Rev. B* **109**, 144421 (2024).

71. Fu, Z. *et al.* Multiple Topological Phases Controlled via Strain in Two-Dimensional Altermagnets. Preprint at https://doi.org/10.48550/arXiv.2507.22474 (2025).

72. Kumar, S. & Kumar, S. Large interfacial contribution to ultrafast THz emission by inverse spin Hall effect in CoFeB/Ta heterostructure. *iScience* **25**, 104718 (2022).

73. Gueckstock, O. *et al.* Terahertz Spin-to-Charge Conversion by Interfacial Skew Scattering in Metallic Bilayers. *Advanced Materials* **33**, 2006281 (2021).

74. Xu, R. *et al.* Terahertz generation via the inverse orbital Rashba-Edelstein effect at the Ni/CuOx interface. *Phys. Rev. Res.* **7**, L012042 (2025).

75. El Hamdi, A. *et al.* Observation of the orbital inverse Rashba–Edelstein effect. *Nat. Phys.* https://doi.org/10.1038/s41567-023-02121-4 (2023) doi:10.1038/s41567-023-02121-4.